\documentclass[12pt]{article}
\input epsf.sty
\newcommand{\omg}{\omega}
\newcommand{\ff}{F}

\newcommand{\ra}{\rangle}
\newcommand{\la}{\langle}

\newcommand{\vp}{\varphi}

\newcommand{\tl}{\wt\lambda}
\newcommand{\hl}{\wh\lambda}

\newcommand{\VV}{{\cal V}}
\newcommand{\BB}{{\cal B}}

\newcommand{\HH}{{\cal H}}

\newcommand{\OO}{{\cal O}}

\newcommand{\EE}{{\cal E}}

\newcommand{\wt}{\widetilde}
\newcommand{\wh}{\widehat}

\newcommand{\NN}{{\cal N}}
\newcommand{\TT}{{\cal T}}

\newcommand{\be}{\begin{equation}}
\newcommand{\ee}{\end{equation}}
\newcommand{\ben}{\begin{eqnarray}\displaystyle}
\newcommand{\een}{\end{eqnarray}}
\newcommand{\refb}[1]{(\ref{#1})}
\newcommand{\p}{\partial}
\newcommand{\sectiono}[1]{\section{#1}\setcounter{equation}{0}}

\begin{document}
{}~
\hfill\vbox{\hbox{hep-th/0207105}
}\break

\vskip .6cm

\centerline{\Large \bf
Time Evolution in Open String Theory
}

\medskip

\vspace*{4.0ex}

\centerline{\large \rm 
Ashoke Sen }

\vspace*{4.0ex}

\centerline{\large \it Harish-Chandra Research
Institute}

\centerline{\large \it  Chhatnag Road, Jhusi,
Allahabad 211019, INDIA}

\centerline {and}
\centerline{\large \it Department of Physics, Penn State University}

\centerline{\large \it University Park,
PA 16802, USA}

\centerline{E-mail: asen@thwgs.cern.ch, sen@mri.ernet.in}

\vspace*{5.0ex}

\centerline{\bf Abstract} \bigskip 

We discuss a general iterative procedure for constructing time dependent
solutions in open string theory describing rolling of a generic tachyon
field away from its maximum. These solutions are characterized by two
parameters labelling the initial position and velocity of the tachyon
field, but one of these parameters can be removed by using time
translation symmetry.  The Wick rotated version of the resulting one
parameter family of inequivalent solutions describes a one parameter
family of boundary conformal field theories, each member of which is
related to the boundary conformal field theory describing the original
D-brane system by a nearly marginal deformation. We apply this technique
to construct a time dependent solution on a D-brane in bosonic string
theory which can be interpreted as the creation of a lower dimensional
brane during the decay of an unstable D-brane. 

\vfill \eject

\baselineskip=16pt

\tableofcontents

\sectiono{Introduction and Summary} \label{s1}

Time dependent solutions in string theory have received attention 
recently, both in closed string theory\cite{closed} and in open 
string 
theory\cite{0202210,0203211,0203265,0204203,0205085,0205098,0206102,0207004}.
(For early studies of open string tachyons see \cite{old}.)
In
particular in 
open string 
theory a 
special class of time 
dependent solutions were constructed which describe the rolling of the 
tachyon field on a non-BPS D-brane\cite{9806155,9809111}, or a 
brane-antibrane pair away from the 
maximum of the potential. Possible applications of these solutions to 
cosmology have been discussed in \cite{cosmo}.\footnote{For earlier 
attempts to 
apply the rolling tachyon solution to cosmology, see \cite{earl}.}

One common feature of the solutions associated with the rolling of an open 
string tachyon is that after removing the trivial 
parameter of the solution associated with time translation, the solutions 
are characterized by one parameter labelling the energy of the
solution.
This is precisely what one
would expect in a scalar field theory
with standard kinetic term, where a general spatially 
homogeneous solution 
will be characterized by the initial position and velocity of the scalar 
field, and one of these two parameters can be removed by using time 
translation invariance which allows us to set either the initial position 
or the initial velocity to zero.\footnote{In a multi-well potential there 
can be more than one
family of inequivalent solutions labelled by the energy.}
However this is somewhat surprising in 
open string field theory where the interaction terms have infinite number 
of derivatives.\footnote{Of course while the analysis of 
\cite{0203211,0203265} generates 
a one parameter family of time dependent solutions, it does not rule out 
the existence of other solutions and so it is still conceivable that the 
full theory has more solutions.}

The analysis of \cite{0203211,0203265} 
was facilitated by the fact that the rolling 
tachyon solution was related by Wick rotation to a boundary conformal 
field theory (BCFT) that was obtained from the original D-brane system 
by an 
{\it exactly marginal} deformation. The deformation parameter was related 
to 
the parameter labelling the inequivalent solutions. We cannot hope that 
this will be true for 
more general tachyonic states which might be present on a generic D-brane 
system, {\it e.g.} tachyons on an intersecting brane 
system\cite{9704006,9703217}.\footnote{Possible application of rolling
tachyon on 
intersecting D-brane system to cosmology has been emphasized in
\cite{inter}.} Thus 
the question arises: is it still possible to construct a one 
parameter family of inequivalent solutions which describe the rolling of a 
generic 
open string tachyon away from its maximum? 

This is the question we address 
in the paper, and show that under certain generic conditions, the answer 
to this question is in the affirmative. In particular once these 
conditions are satisfied, we have a one parameter family of inequivalent 
solutions, which are related by Wick rotation to a one parameter family of 
euclidean BCFT's. However, these BCFT's are not related to each other by a 
a marginal deformation. Instead, each of them is related to the BCFT 
describing the original unstable D-brane system by a nearly marginal 
deformation. The nearly marginal operator, however, is 
different for 
different solutions.

Unfortunately, although our analysis establishes the existence of a one 
parameter family of rolling tachyon solutions for a generic open string 
tachyon, unlike in the cases discussed in \cite{0203211,0203265},
generically the 
deformed BCFT's are 
not exactly solvable. As a result, we cannot give an explicit construction 
of 
the energy-momentum tensor by Wick rotating the boundary state of the 
deformed theory. Nevertheless, the success of this approach in generating 
explicit
solution in special cases where the deformation is exactly 
marginal\cite{0203211,0203265}, as well as in generating general 
class of solutions in 
$p$-adic string theory\cite{mo-zw} leads us to believe that this approach 
has a more 
general validity.


In section \ref{s2} we 
illustrate our 
method of constructing the rolling tachyon solution by studying the 
example of a 
scalar field theory with standard kinetic term and potential $V(\phi)$ 
with a maximum at $\phi=0$. The idea is to relate the 
rolling tachyon solution via Wick rotation to the equation of motion for a 
scalar field with potential $-V(\phi)$. We can now solve for the periodic 
motion of this scalar field around $\phi=0$ using perturbation 
theory where we let the period of oscillation depend on the amplitude, and 
solve for the period and the orbit as a perturbation series in the 
amplitude\cite{golds}. Once we have obtained a solution this way we can 
inverse Wick 
rotate the solution to find a rolling tachyon solution. The amplitude of 
oscillation in the Wick rotated theory labels the initial
value of the 
rolling tachyon.

In section \ref{s3} we show how under certain conditions this procedure
can be generalized to open string field theory to construct rolling
tachyon solutions associated with a generic tachyon field. 
We also generalize this construction to the case where multiple tachyons
roll simultaneously beginning with arbitrary initial position and
velocity. We reinterpret
this construction in section \ref{s4} in terms of boundary conformal field
theory. In particular we show that in the Wick rotated theory, the family
of solutions labelled by the initial position of the tachyon field 
correspond 
to a family of BCFT's, and each member of this family
is
related to the BCFT describing the original D-brane by a nearly marginal
deformation. 

In section \ref{s5} we consider a specific example -- that of
a D-$p$-brane in bosonic string theory with one direction compactified on 
a circle of radius $R$,
-- and consider the rolling of the $(p-1,1)$ dimensional tachyon, obtained
by taking the first momentum mode of the $(p,1)$ dimensional tachyon along
the circle. We show how in the Wick rotated theory we can construct a one
parameter family of boundary conformal field theories, each related to the
original D-$p$-brane by a nearly marginal deformation, and how the inverse
Wick rotation of these BCFT's generate the family of rolling tachyon
solutions.  Unfortunately these BCFT's are not exactly solvable and hence
we cannot compute the analytic expressions for the energy-momentum tensor 
associated with these solutions.
However at a particular value of the radius, $R=\sqrt 2$, the family of 
BCFT's become related by a marginal deformation and are exactly solvable. 
As a result we can compute the time evolution of the energy momentum 
tensor explicitly by taking the inverse Wick rotation of the boundary 
states associated with these BCFT's. We find that during the course of 
time evolution there is non-trivial flow of energy density along the 
compact direction, and at a certain finite value of 
time, the energy density at a particular location on this circle
blows up. More specifically, the energy density as a function of the
coordinate $y$ along the compact circle approaches a delta function
singularity as we approach this critical time.
As we pass this  critical time, there is an apparent loss of energy
density at the location of the singularity, with the total amount of
energy lost being equal to the total initial energy of the system. Thus
just after the critical time the energy density averages to zero. If we
naively continue the
formula for energy density beyond this critical time, the energy density
gradually evolves to zero
everywhere. We suggest a natural interpretation of this as the creation of
a codimension one lump from the decay of the original brane.

In section \ref{s6} we generalize the construction to the case of 
superstring theory. In particular we consider rolling tachyon solution 
that should describe the creation of a D-$(p-1)$ -- $\bar{\rm D}$-$(p-1)$ 
pair due to the decay of a non-BPS D-$p$-brane wrapped on a circle. As in
the case of bosonic 
string theory, we can construct 
this solution in the Wick rotated theory as a nearly marginal deformation 
of the original D-$p$-brane conformal field theory. However, this BCFT is 
not 
exactly solvable, and hence we cannot explicitly construct the 
space-time dependent energy 
momentum tensor associated with this solution.
In section \ref{s7} we discuss 
possible generalization of this method to the study of closed string
tachyon condensation. Although 
part of the argument can be generalized to the case of closed strings, the 
main obstacle to finding the solution in closed string theory arises 
due to our inability to solve the equations of motion of the graviton and 
the dilaton field in the background of rolling tachyon.
We conclude in section \ref{s8} with a few remarks.

\sectiono{Example in Scalar Field Theory} \label{s2}

In this section we shall discuss the general method for constructing time 
dependent solutions in the context of a scalar field theory, which we 
shall 
generalize to string field theory in the later sections. We begin 
with the action of a scalar field $\phi$ in $p+1$ dimensions with standard 
kinetic term and potential $V(\phi)$:
\be \label{e1}
S = - \int d^{p+1} x [ \p^\mu\phi \p_\mu\phi + V(\phi)]\, .
\ee
We shall further assume that $V(\phi)$ has a maximum at $\phi=0$, with
\be \label{e2}
V''(0) = -m^2\, .
\ee
We want to study time dependent solution of this equation determined by a
given initial condition. For simplicity let us consider spatially
homogeneous field configurations. In this case the solution is
characterized by two parameters, -- the initial position and velocity of
$\phi$. For definiteness we shall further restrict to solution with
total energy $E< V(0)$. In this case $\p_0\phi$ vanishes at some instant
of time when $V(\phi)=E$. We shall take this to be the origin of $x^0$. 
Thus the solution is now characterized by only one parameter $\lambda$, -- 
the value of $\phi$ at $x^0=0$. We shall be interested in the solution 
where $\lambda$ is small but not infinitesimal.

The equation of motion is:
\be \label{e3}
\p_0^2 \phi + V'(\phi) = 0\, .
\ee
It is of course straightforward to (numerically) integrate this equation, 
but 
this cannot be generilized to string field theory where the equations of 
motion contain infinite number of derivatives. 
We shall follow an indirect method that can be generalized in the context 
of 
string field theory. The basic idea is the same as the one followed in 
\cite{0203211,0203265}, namely we make a Wick rotation $x^0=i x$, and 
write eq.\refb{e3} as
\be \label{e4}
\p_x^2 \phi - V'(\phi) = 0\, .
\ee
If $\phi=f(x)$ is a solution of eq.\refb{e4}, then $\phi=f(-i x^0)$ will 
be a solution to \refb{e3}. Thus the aim is now to solve eq.\refb{e4} 
with the boundary condition $\phi=\lambda$, $\p_x\phi=0$ at $x=0$. 
This clearly can be thought of as a motion of a particle in potential 
$-V(\phi)$. Since $V(\phi)$ has a maximum at $\phi=0$, $-V(\phi)$ has a 
minimum at $\phi=0$, and for small $\lambda$, solution to the equation of 
motion 
will oscillate around zero. 
The period of oscillaton $T\equiv 2\pi/\omg$ is in general a function of 
$\lambda$.\footnote{An analytic expression of 
$T$ is
given by
$$ T = \sqrt 2 \, \int_{\phi_1}^{\phi_2} {d\phi
 \over \sqrt{V(\phi)-V(\lambda)}}\, ,
$$ 
where $\phi_1$ and $\phi_2$ are turning points where 
$V(\lambda)=V(\phi_1)=V(\phi_2)$. The initial value $\lambda$ of $\phi$ 
can be identified with either $\phi_1$ or $\phi_2$. However, we shall not 
use this formula.} Thus the 
solution can be expanded as
\be \label{e9}
\phi(x; \lambda) = \sum_{n=0}^\infty a_n \cos(n\omg x)\, .
\ee

We shall now solve \refb{e4} using perturbation 
theory\cite{golds}.
If we write
\be \label{e5a}
V(\phi) = -{1\over 2} m^2 \phi^2 +  V_{int}(\phi)\, ,
\ee
then \refb{e4} takes the form
\be \label{e5b}
(\p_x^2 + m^2) \phi -  V_{int}'(\phi) = 0\, .
\ee 
$ V_{int}(\phi)$ is of order $\phi^n$ with $n\ge 3$. Thus
for small $\lambda$ the solution behaves as
\be \label{e10}
\phi = 
\lambda\cos(mx) + \OO(\lambda^2)\, . 
\ee
Comparing this with \refb{e9} we get
\be \label{e11}
a_1 = \lambda + \OO(\lambda^2), \qquad
\omg = m + \OO(\lambda), \qquad 
a_n = \OO(\lambda^2) \quad \hbox{for} \quad n=0, n\ge 2\, .
\ee
{}From this we see that we can trade in the parameter $\lambda$ for 
the coefficient $a_1\equiv\hl$, and rewrite \refb{e11} as 
\be \label{e8}
a_1=\hl,  \qquad
\omg = m + \OO(\hl), \qquad
a_n = \OO(\hl^2) \quad \hbox{for} \quad n=0, n\ge 2\, .
\ee
We now substitute \refb{e9} into eq.\refb{e5b} to get
\be \label{e12}
\sum_{n=0}^\infty \Big({n^2\omg^2}-m^2\Big) \, a_n \cos(n\omg x)
= - V_{int}'\Big( \sum_{n=0}^\infty \, a_n \cos(n\omg x)\Big)\, .
\ee
The right hand side of this equation contains terms quadratic and higher 
order in $a_n$. Thus we can solve the equations iteratively as follows. 
The zeroeth order approximation is taken to be
\be \label{e13}
\omg=m, \qquad a_n=0 \quad \hbox{for} \quad n=0, \quad n\ge 2\, .
\ee
Also $a_1$ is set equal to $\hl$ to all orders.
We 
substitute the $k$-th order results for the $a_n$'s and $\omg$ on the 
right 
hand side of \refb{e12} to compute the coefficient of $\cos(n\omg x)$ 
for every $n$. 
Comparing this with the coefficient of $\cos(n\omg x)$ on the left hand 
side, 
we determine the $(k+1)$th order values of the $a_n$'s for $n=0$ and $n\ge 
2$. On the other hand equating the coefficient of the $\cos(\omg x)$ 
terms on 
both sides, we determine the value of $\omg$ to $(k+1)$th 
order.\footnote{Actually this determines $\omg$ to order $\hl^k$, but 
since the coefficients $a_n$ are of order $\hl$ or higher, this determines 
the solution to order $\hl^{k+1}$.} This is 
possible since $a_1$ is set equal to $\hl$, and we are determining all the 
coefficients in terms of $\hl$.

This gives a solution to eq.\refb{e4}. Given this, we can now generate a 
solution to eq.\refb{e3} by making the substitution $x=-ix^0$. This gives
\be \label{e14}
\phi(x^0) = \sum_{n=0}^\infty a_n(\hl) \cosh(n\omg(\hl) x^0)\, ,
\ee
where the coefficients $a_n$ are the same as the ones determined by the 
previous method. This gives a family of solutions characterized by one 
parameter $\hl$. $\hl$ determines the initial value $\lambda$ of $\phi$, 
with the precise relation between $\hl$ and $\lambda$ being given by:
\be \label{e15}
\lambda = \phi(0) = \sum_{n=0}^\infty a_n(\hl)\, .
\ee

\begin{figure}[!ht]
\leavevmode
\begin{center}
\epsfysize=5cm
\epsfbox{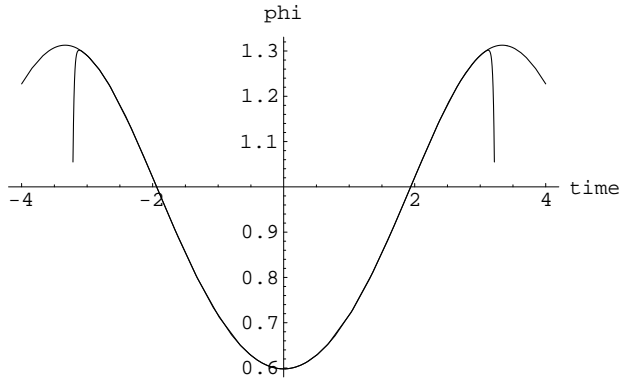}
\end{center}
\caption{The plot of a particular time dependent solution in a scalar 
field theory with potential $-\phi^2/2 + \phi^3/3$ for the choice
$\hl=.5$. This gives $\phi(x^0=0)\simeq 0.598122$. The top curve is the 
result of direct numerical integration with this boundary condition, while
the bottom curve is obtained 
by perturbative techniques discussed here with 60 harmonics. As we can see 
the two curves 
coincide for $-\tau/2 < x < \tau/2$ where $\tau$ is the period 
of oscillation, but the perturbation theory breaks down beyond this 
range.} \label{f1}
\end{figure}

For example, if we take the $\phi^3$ field theory with potential $-{1\over
2}\phi^2 + {1\over 3}\phi^3$, then $V'_{int}(\phi) = \phi^2$. Using the
first order solution $\phi(x^0)=\hl\cos(\omega x^0)$ on the right hand
side of \refb{e12}, we get the solution to second order in $\hl$: 
\be \label{esecond}
a_1 = \hl, \qquad a_0 \simeq {1\over 2} \hl^2, \qquad a_2 \simeq -{1\over
6}
\hl^2, \qquad \omega \simeq 1\, .
\ee
Note that $\omega$ remains constant at 1 since the Fourier expansion of
$(\hl\cos x)^2$ does not contain a term proportional to $\cos x$.
Substituting \refb{esecond} on the right hand side of \refb{e12} and
comparing coefficients of $\cos(\omega x)$ on both sides we get the
equation for $\omega$ at next order:
\be \label{eyyy1}
(\omega^2-1) \hl \simeq -{5\over 6} \hl^3 \qquad \to \qquad \omega^2
\simeq 1
-{5\over 6} \hl^2\, .
\ee
The coefficients $a_n$ to order $\hl^3$ are given by:
\be \label{eyyy2}
a_1 = \hl, \qquad a_0 \simeq {1\over 2} \hl^2, \qquad a_2 \simeq -{1\over
6}
\hl^2, \qquad a_3 \simeq {1\over 48} \hl^3\, .
\ee
The explicit expression for the solution becomes complicated at higher
order, but we can evaluate it numerically. 
As illustrated in Fig.\ref{f1},
explicit numerical analysis in this $\phi^3$ field theory shows that this 
procedure generates the time dependent solution in this theory, describing 
rolling of $\phi$ from the maximum towards the local minimum of 
$V(\phi)$, quite accurately for half a
period of oscillation in either direction of $x^0$. 
Beyond this range the expansion \refb{e14} 
diverges and we must take recourse to analytic continuation in order to 
study the solution. 

\sectiono{Solution in String Field Theory} \label{s3}

We now consider some specific D-brane system in bosonic string theory with 
a tachyonic mode. Associated with the zero momentum tachyon state, there 
is a boundary operator $V_T$ of dimension $h<1$. In $\alpha'=1$ unit, the 
associated tachyon 
field has mass$^2$
\be \label{e2.1}
(h-1) \equiv -m^2\, .
\ee
If we displace the tachyon field $T$ a distance $\lambda$ away from its 
maximum, and let the system evolve in time, then to leading order in 
$\lambda$ the solution is given by 
$\lambda\cosh(mx^0)$.  In the Wick rotated theory this corresponds to 
$\lambda\cos(mx)$.
The corresponding solution of the linearizerd 
equations 
of motion in string field theory is given by
\be \label{e2.2}
\lambda\, \cos(mX(0)) \, V_T(0)c_1|0\ra\, .
\ee
This is a BRST invariant state of $L_0$ eigenvalue 0.

The full string field theory equation of motion is given by:
\be \label{e2.2a}
Q_B |\Psi\ra + |\Psi*\Psi\ra = 0\, ,
\ee
where $|\Psi\ra$ is an open string state of ghost number 1, and
$Q_B$ is the BRST 
charge. We shall try to find a solution to eq.\refb{e2.2a} in the Siegel 
gauge
\be \label{e2.2b}
b_0|\Psi\ra = 0\, ,
\ee
where the equation takes the form:
\be \label{e2.2c}
L_0|\Psi\ra = -b_0 |\Psi*\Psi\ra\, .
\ee
Our goal will be to first construct a solution to \refb{e2.2c} that
extends \refb{e2.2} to higher orders in $\lambda$, and then carry
out
an inverse Wick rotation to  construct a time dependent solution to
eq.\refb{e2.2c} 
describing the rolling of the tachyon away from its maximum.
We begin by defining:
\be \label{e2.3}
|\phi_0\ra = \cos(\omg X(0)) \, V_T(0)\, c_1|0\ra\, ,
\ee 
where $\omg$ is a number that will be determined 
below.
We choose a set of linearly independent basis states $|\phi_r\ra$ ($0\le 
r < \infty$) in a
subspace $\HH_1$ of first quantized open string states, 
satisfying the following conditions:
\begin{enumerate}
\item Each $|\phi_r\ra$ has ghost number 1.
\item Each $|\phi_r\ra$
is symmetric under $X\to -X$, and carries $x$-momentum which 
is an
integer multiple of $\omg$.
\item All $|\phi_r\ra$ for $r\ge 1$ are orthogonal to $c_0|\phi_0\ra$.
\end{enumerate}
$\HH_1$ by definition is the subspace spanned by the basis states
$\{|\phi_r\ra, r\ge 0\}$ satisfying these properties, and
we shall look for a solution $|\Psi\ra$ of 
eq.\refb{e2.2c} in this subspace. We shall also require that when we 
express the solution $|\Psi\ra$ as a linear combination of the basis 
states $|\phi_r\ra$, the 
coefficient of $|\phi_0\ra$ in $|\Psi\ra$ is given by $\hl$; this in fact 
will be the definition of the parameter $\hl$. For later use, we shall 
define:
\be \label{edefpri}
|\chi\ra' = |\chi\ra - \NN^{-1} \la\phi_0|\chi\ra \, c_0|\phi_0\ra, \qquad 
\NN \equiv \la \phi_0|c_0|\phi_0\ra\, ,
\ee
for any state of $|\chi\ra$ ghost number 2. Physically $|\chi\ra'$ gives 
$|\chi\ra$ with its component along $c_0|\phi_0\ra$ removed.

The construction of the solution $|\Psi_{(n)}\ra$ to order $\hl^n$
proceeds as follows.
We define
\be \label{e2.4}
|\Psi_1\ra = \hl |\phi_0\ra\, ,
\ee
and  $|\Psi_p\ra$ iteratively by using the relation:
\be \label{e2.5}
|\Psi_p\ra = -{b_0\over L_0} (|\Psi_{p-1}*\Psi_{p-1}\ra)' + \hl 
|\phi_0\ra\, .
\ee
We repeat \refb{e2.5} $(n-1)$-times 
to compute $|\Psi_n\ra$. 
If we define
\be \label{edeffn}
f_n(\omg,\hl) = \NN^{-1} \, \la \phi_0| \Psi_{n-1} * \Psi_{n-1}\ra\, ,
\ee
then we have
\be \label{e2.6}
(|\Psi_{n-1}*\Psi_{n-1}\ra)' = |\Psi_{n-1}*\Psi_{n-1}\ra - f_n(\omg,\hl) 
c_0 |\phi_0\ra\, .
\ee
We now define $\omg_n$ as the solution of the equation:
\be \label{e2.9}
\hl (\omg_n^2 - m^2)+
f_n(\omg_n,\hl) = 0\, ,
\ee
and take the $n$-th order solution $|\Psi_{(n)}\ra$ to be
\be \label{e2.11a}
|\Psi_{(n)}\ra = |\Psi_n\ra_{\omg=\omg_n}\, .
\ee
Note that $\omg$ at $n$th order is determined {\it after we have 
computed $f_n(\omg,\hl)$ by repeating the iterative procedure $(n-1)$ 
times}. 
Since $f_n(\omg,\hl)$ is of order $\hl^2$, \refb{e2.9} gives 
$\omg_n=m+\OO(\hl)$.\footnote{Actually from the definition of $\phi_0$ 
and 
conservation of $X$-momentum, it follows 
that $|\phi_0*\phi_0\ra$ contains states with $X$-momentum $\pm 2\omg$ 
or $0$, but does not contain a state of momentum $\pm\omg$. Thus 
$|\phi_0*\phi_0\ra$ does not contain any component along $c_0|\phi_0\ra$. 
As a result the order $\hl^2$ contribution to $f_n(\omega,\hl)$ vanishes,
and 
$f_n(\omega,\hl)=\OO(\hl^3)$. Thus $\omg=m+\OO(\hl^2)$. Generalising this 
argument we can see that 
$f_n(\omega,\hl)$ must be an odd function of $\hl$, and hence $\omg$ must
be 
an even function of $\hl$.}

We shall now show that $|\Psi_{(n)}\ra$ determined this way 
satisfies:
\be \label{e2.11}
L_0 |\Psi_{(n)}\ra = - b_0 |\Psi_{(n)}*\Psi_{(n)}\ra + 
\OO(\hl^{n+1})\, ,
\ee
and hence satisfies the string field theory equations of motion to order
$\hl^n$.
To do this, we
first note that \refb{e2.5}, \refb{e2.6} for $p=n$ gives
\be \label{e2.7}
L_0|\Psi_n\ra = -b_0 |\Psi_{n-1}*\Psi_{n-1}\ra +   
f_n(\omg,\hl)|\phi_0\ra  + \hl (\omg^2 - m^2) |\phi_0\ra\, ,
\ee
where we have used the fact that
\be \label{e2.8}
L_0 |\phi_0\ra = (\omg^2 - m^2) |\phi_0\ra\, .
\ee
Eq.\refb{e2.9}, \refb{e2.7} now give 
\be \label{e2.10} 
[L_0 |\Psi_n\ra + b_0 |\Psi_{n-1}*\Psi_{n-1}\ra]_{\omg=\omg_n}=0\, .
\ee 
We shall now show that 
\be \label{e2.12}
|\Psi_p\ra = |\Psi_{p-1}\ra + \OO(\hl^p)\, ,
\ee 
for any $p\le n$.
In that case if we replace $|\Psi_{n-1}\ra$ on 
the right hand side of \refb{e2.10} by $|\Psi_n\ra$, the net error will be 
of order $\hl^{n+1}$, since $\Psi_{n}$ itself is order $\hl$. 
This, together with \refb{e2.11a}, would establish \refb{e2.11}.

In order to establish \refb{e2.12} we use proof by induction.
We use \refb{e2.5} to get:
\be \label{e2.13}
|\Psi_{p+1}\ra - |\Psi_{p}\ra 
= -{b_0\over L_0} (|\Psi_{p}*\Psi_{p}\ra - 
|\Psi_{p-1}*\Psi_{p-1}\ra)'\, .
\ee
Now suppose that \refb{e2.12} holds for some value of $p$. In that case 
the 
right hand side of \refb{e2.13} 
is of 
order 
$\hl^{p+1}$, since
$|\Psi_{p-1}\ra$ itself is of order $\hl$. Since \refb{e2.12} holds 
for $p=1$, this proves \refb{e2.12} for 
all 
$p\le n$.

Note however that for the above argument to go through, we need to ensure 
that $b_0(|\Psi_{p}*\Psi_{p}\ra -
|\Psi_{p-1}*\Psi_{p-1}\ra)'$ does not have a state with $L_0$ eigenvalue 
of order $\hl$, since this will give an additional power of $\hl$ in the 
denominator from the operation of $1/L_0$ and will destroy the counting of 
powers of $\hl$ that we have used. It is precisely due to this reason that 
we have removed the component of $|\Psi_{p}*\Psi_{p}\ra$ along 
$c_0|\phi_0\ra$ in this expression, since $c_0|\phi_0\ra$ has $L_0$ 
eigenvalue 
$(\omg^2 - m^2)\sim \hl$. For our analysis to hold it is important that
besides $|\phi_0\ra$,  
there are no other states appearing in the computation of 
$|\Psi_{p}*\Psi_{p}\ra$, which has $L_0$ eigenvalue of order $\hl$ or 
less.

This requirement can be reformulated purely in terms of matter sector 
vertex operators as follows.
The possible problematic states appearing in the expression for
$|\Psi_{p}*\Psi_{p}\ra$ are ghost number 2 states which are not
annihilated by $b_0$, and has $L_0$ eigenvalue 0 in the $\hl\to 0$ limit.  
Such states are of the form: 
\be \label{ef1} 
O(0)\, c_0 c_1 |0\ra\, , 
\ee
where $O$ is a matter sector vertex operator whose conformal weight
approaches 1 as $\hl\to 0$. As long as the theory does not contain any
operator of this type which can appear in the operator product of
$\cos(\omg X) V_T$ with itself (other than $\cos(\omg X) V_T$ itself,
whose effect has already been taken into account) the analysis given above
remains valid.  Matter operators of the form $\p X$ of dimension 1, which 
could cause potential problems, do not do so since the $X\to -X$ symmetry 
prevents their appearance in 
the operator product of $\cos(\omg X) V_T$ with itself.

The situation simplifies if not even $c_0|\phi_0\ra$ appears in the 
computation of $|\Psi_p*\Psi_p\ra$ for $\omega=m$.
In this case we see from \refb{edeffn} that
$f_n(m,\hl)$ vanishes identically for arbitrary $\hl$. Thus 
eq.\refb{e2.9} can be solved by setting $\omg=m$. 
The cases analysed in refs.\cite{0203211,0203265} are of this type. 

Once we have obtained the solution in the Wick rotated theory, we can 
perform an inverse Wick rotation $x=-ix^0$ to obtain a time dependent 
solution in the Lorentzian theory. This amounts to replacing $\cos(\omg 
X)$ by $\cosh(\omg X^0)$, and the oscillator $\alpha_m$ of $X$ by 
$-i\alpha^0_m$ of $-iX^0$. Performing these operations on $\Psi_{(n)}$ we 
can generate a time dependent solution of the equations of motion to order 
$\hl^n$. The constant $\hl$ labelling the solution parametrizes the 
initial 
condition on the tachyon field. Of course, we expect the  series expansion 
to be valid only for a limited range of $x^0$, and for $x^0$ beyond this 
range we need to obtain the result via analytic continuation.

Since the discussion so far has been somewhat abstract, we shall now give 
explicit form of the solution to order $\hl^3$. We have:
\be \label{eexp1}
|\Psi_1\ra = \hl |\phi_0\ra\, ,
\ee
\be \label{eexp2}
|\Psi_2\ra = \hl |\phi_0\ra - \hl^2 \, {b_0\over L_0} \, 
|\phi_0*\phi_0\ra\, .
\ee
No subtraction is needed to this order since $|\phi_0*\phi_0\ra$ does not 
have a component along $c_0|\phi_0\ra$ due to $X$-momentum conservation. 
To next order\footnote{We could ignore the $\OO(\hl^4)$ term in 
$|\Psi_3\ra$, coming from the $*$-product of ${b_0\over L_0} \,
|\phi_0*\phi_0\ra$ with itself, in computing the solution to order
$\hl^3$.} \ben \label{eexp3}
|\Psi_3\ra &=& \hl |\phi_0\ra - \hl^2 \, {b_0\over L_0} |\phi_0 * 
\phi_0\ra \nonumber \\ 
&& + {b_0\over L_0} \bigg( \hl^3 |\phi_0\ra * {b_0\over L_0} \,
|\phi_0*\phi_0\ra
+ \hl^3 {b_0\over L_0} \, |\phi_0*\phi_0\ra * |\phi_0\ra + f_3(\omg,\hl)
c_0 
|\phi_0\ra \bigg) \nonumber \\ \cr
&& + \OO(\hl^4)
\een
where
\ben \label{eexp4}
f_3(\omg,\hl) &=& -\NN^{-1} \hl^3 \la \phi_0| \bigg( |\phi_0\ra * 
{b_0\over 
L_0} \, |\phi_0*\phi_0\ra + {b_0\over L_0} \, |\phi_0*\phi_0\ra * 
|\phi_0\ra \bigg)\nonumber \\
&=& -2\, \NN^{-1} \hl^3 \la \phi_0 * \phi_0 | {b_0\over L_0} |\phi_0 * 
\phi_0\ra
\, .
\een
This can be related to the four point amplitude $A^{(4)}$ in string field 
theory involving four 
external states $|\phi_0\ra$. Taking into account the sum over $s$, $t$ 
and $u$ channel diagrams, and a factor of 2 coming from each vertex since 
the three point coupling in string field theory is accompanied by a factor 
of $1/3$ instead of $1/6$, we have\footnote{Throughout this paper 
$A^{(4)}$ will 
denote the amplitude in the euclidean string (field) theory.}
\be \label{eexp5}
A^{(4)}(\omg) = 12 \, \la \phi_0 * \phi_0 | {b_0\over L_0} |\phi_0 * 
\phi_0\ra \, .
\ee
Thus \refb{eexp4} can be written as
\be \label{eexp6}
f_3(\omg,\hl) = - {1\over 6} \NN^{-1} \hl^3 A^{(4)}(\omg)\, .
\ee
$\omg_3$ is then the solution to the equation
\be \label{eexp7}
(\omg_3^2 - m^2) = {1\over 6} \NN^{-1} \hl^2 A^{(4)}(\omg_3)\, .
\ee
To leading order we can set $\omg_3=m$ on the right hand side of 
\refb{eexp7} and get
\be \label{eexp8}
\omg_3 = \sqrt{ m^2 + {1\over 6} \NN^{-1} \hl^2 A^{(4)}(m)}\, .
\ee
$|\Psi_3\ra$ given in \refb{eexp3}, evaluated at $\omg=\omg_3$, then 
gives the solution $|\Psi_{(3)}\ra$ to order $\hl^3$.

This concludes our discussion on generating the time dependent solution of 
string field theory, describing the rolling of a tachyon away from its 
maximum. We shall end this section by mentioning three generalizations of
this analysis:
\begin{itemize}
\item We have discussed the case where the tachyon begins rolling from an 
initial configuration where its time derivative vanishes and the field is 
displaced from its maximum. In this case the total energy of the system is 
less than that at the maximum. We could also consider the case where we 
begin with a configuration where the tachyon field is at its maximum and 
has a small velocity. The leading order solution in this case 
is proportional to $\sinh(\omg x^0)$ with $\omg=m$. As discussed in 
\cite{0203211,0203265}, 
the full solution in this case can be found by making the replacement:
\be \label{ereplace}
x^0 \to x^0 + i \pi/(2\omg), \qquad  \hl \to - i \hl\, ,
\ee
in the solution we have obtained earlier.
\item 
The method discussed here 
can also be used to generate time dependent solutions describing 
oscillation of positive 
mass$^2$ fields about the minimum of their potential. In this case there 
is no need for Wick rotation. For a field $\phi$ of mass $M$, we take the 
first 
order solution to be
\be \label{ep1}
\hl\, \cos(\omg X^0(0)) \, V_\phi(0)c_1|0\ra\, ,
\ee
where $V_\phi$ is the vertex operator for the zero momentum scalar. 
At leading order, $\omg=M$.
We can now follow the iterative procedure described in this section to 
generate the 
solution to arbitrary power of $\hl$, and determine $\omg$ as a function 
of $\hl$ to that order. The $\hl$ dependence of $\omg$ will represent 
the anharmonicity of the oscillator due to the interaction terms in the 
string field theory action.

\item One could consider generalizing the construction to the case where
many tachyon fields roll simultaneously. In general, if we have $n$ scalar
fields then for a two derivative action we have $2n$ initial conditions.
We shall now show that even in string field theory we can construct a $2n$
parameter family of solutions describing the rolling of $n$ tachyons.
To leading
order the solution in the Wick rotated theory is:
\be \label{egena1}
|\Psi\ra = \sum_{i=1}^n \hl^{(i)} |\phi_0^{(i)}\ra\, ,
\ee
where
\be \label{egena2}
|\phi_0^{(i)}\ra = \cos(\omega^{(i)} X(0)+\vp^{(i)}) V_T^{(i)}(0)
c_1|0\ra\, .
\ee
$V_T^{(i)}$ is the vertex operator of the $i$th tachyon with
mass$^2=-m_{(i)}^2$, $\{\hl^{(i)}\}$ and $\{ \vp^{(i)}\}$ 
are the $2n$ parameters labelling the solution, and to leading order
$\omega^{(i)}=m^{(i)}$. We
shall assume that the $m^{(i)}$'s are incomensurate.
We
now generate the higher order solutions by iterating
\refb{egena1} as in the case of a single tachyon field, keeping
the coefficient of $|\phi^{(i)}_0\ra$ held fixed at $\hl^{(i)}$ to all
orders. The analog
of $|\chi\ra'$ for a state $|\chi\ra$ of ghost number 2 is defined as the
projection of $|\chi\ra$ in the subspace orthogonal to the one spanned by
the states $|\phi^{(i)}_0\ra$. Thus comparison of the coefficients of
$c_0|\phi^{(i)}_0\ra$ terms in the equation of motion determines
$\omega^{(i)}$ as a function of the $\hl^{(j)}$'s, whereas the
comparison of the other terms in the equation of motion determine the
component of $|\Psi\ra$ in the subspace orthogonal to the one spanned by
the $c_0|\phi^{(i)}_0\ra$'s. At the end of the computation we need to make
the replacement $X\to -iX^0$, and $\vp^{(i)}\to -i \theta^{(i)}$ where
$\theta^{(i)}$ are taken to be real parameters.

For this procedure to work we need to ensure that the operator product of
the operators $\cos(\omega^{(i)} X + \vp^{(i)}) V_T^{(i)}$ in the matter
sector does not generate another operator of dimension $=\OO(\hl^{(i)})$
outside the set $\{\cos(\omega^{(i)} X + \vp^{(i)})V_T^{(i)}\}$. The
generic dangerous operators are $\sin(\omega^{(i)} X +
\vp^{(i)})V_T^{(i)}$ and $\p
X$. The fact that $\sin(\omega^{(i)} X + \vp^{(i)})V_T^{(i)}$ is not
generated can be seen as follows.
If in a correlation
function we have one insertion of $\sin(\omega^{(i)} X + \vp^{(i)})$ and
multiple insertions of $\cos(\omega^{(j)} X + \vp^{(j)})$ for different
$j$, then we can evaluate this by writing the sines and cosines as sum of
exponentials. 
As long as the $\omega^{(j)}$'s are incomensurate,\footnote{This
restriction on $\omega^{(j)}$ means that the points in the parameter space
where perturbation theory fails are dense in the full parameter space, but
the situation here is no different from that in perturbation theory in
ordinary Hamiltonian system.} a correlator involving
$\exp(\pm i (\omega^{(j)}X + \vp^{(j)}))$ will be non-zero only if the
$\pm\omega^{(j)} X$ in the exponent cancel pairwise. As a result
the accompanying $\pm\vp^{(j)}$ also cancel pairwise. It is now easy to
see that for each term in the correlator involving the
$e^{i(\omega^{(i)} x + \vp^{(i)})}$ in $\sin(\omega^{(i)} x + \vp^{(i)})$,
there is an equal and opposite contribution involving the 
$e^{-i(\omega^{(i)} x + \vp^{(i)})}$ term in $\sin(\omega^{(i)} x +
\vp^{(i)})$, obtained by reversing the signs of the exponents of all the
other terms in the correlator.\footnote{There are many ways to pair terms
so that they cancel; here we only mention one of them.}
These two terms cancel. Thus such a
correlator vanishes, implying in turn that the operator product of
$\cos(\omega^{(j)} X + \vp^{(j)})$ does not contain a term proportional to
$\sin(\omega^{(i)} X + \vp^{(i)})$.

To see the absence of $\p X$ in the operator product of $\cos(\omega^{(j)}
X + \vp^{(j)})$, we consider a correlation function with one insertion of
$\p X$ and multiple insertions of various factors of $\cos(\omega^{(j)} X
+ \vp^{(j)})$ on the boundary of the world-sheet. Since $\p X$ can be
thought of as the restriction of a bulk holomorphic current to the
boundary, this contribution can be expressed as a sum of poles of $\p X$
at the location of the various insertions of $\cos(\omega^{(j)} X +
\vp^{(j)})$. The residue of the pole at the insertion of
$\cos(\omega^{(k)} X + \vp^{(k)})$ is given by  
$\omega^{(k)}\sin(\omega^{(k)} X +
\vp^{(k)})$.
Thus
\ben \label{epp1}
&&\bigg\langle \p X(z) \prod_{s=1}^N  \cos(\omega^{(j_s)}
X(t_s)  + \vp^{(j_s)})\bigg\rangle \nonumber \\
&\propto&\sum_{r=1}^N {\omega^{(j_r)}\over (z-t_r)} \bigg\langle
\sin(\omega^{(j_r)}
X(t_r) + \vp^{(j_r)})
\prod_{s\ne r}  \cos(\omega^{(j_s)}
X(t_s) + \vp^{(j_s)})\bigg\rangle \, ,
\nonumber \\
&& \qquad 1\le j_s \le n   
\een
The result vanishes by previous
argument. This, in turn, shows that the correlator of $\p X$ with a set of
$\cos(\omega^{(j)} X + \vp^{(j)})$ vanishes and hence we do not generate
$\p X$ in the operator product of $\cos(\omega^{(j)} X + \vp^{(j)})$.

Of course there may be other special dimension 1 operators which may be
generated in the operator product of $\cos(\omega^{(j)} X + \vp^{(j)})
V_T^{(j)}$. In that case the perturbative procedure outlined here for
generating solutions of string field theory equations of motion breaks
down.

\end{itemize}

\sectiono{Boundary Conformal Field Theory Description} \label{s4}

In this section we shall describe the BCFT associated with the solutions
constructed in the previous section. Since in the euclidean theory the
final solution is obtained by iterating an initial solution of the form
\be \label{e3.1} 
\hl \cos(\omg X(0)) \, V_T(0)\, c_1|0\ra\, , 
\ee 
we
might expect that this corresponds to a BCFT that is obtained by
perturbing the original BCFT by a boundary term
\be \label{e4.1} 
\tl \, \int dt
\cos(\omg X(t))\, V_T(t)\, , 
\ee 
where $\tl=\hl+\OO(\hl^2)$ and $t$ is a parameter labelling the boundary
of the world-sheet. In order for \refb{e4.1} to describe
a BCFT the $\beta$-function of the theory must vanish. Since $\cos(\omg
X)\, V_T$ has conformal weight $(\omg^2 + h)=(\omg^2-m^2+1)$, the
$\beta$ function for the coupling $\tl$ has the form: 
\be \label{e4.2}
\beta_{\tl}=(\omg^2-m^2)\tl + g(\omg, \tl)\, , 
\ee 
where
$g(\omg,\tl)$ represents higher order (in $\tl$) contribution to the
$\beta$-function. Thus the vanishing of the $\beta$ function 
requires:\footnote{See \cite{zam} for a discussion on this. For
application of the $\beta$-function equation to the study of tachyon
condensation, see \cite{bsft,0010247}.} 
\be \label{e4.3} 
(\omg^2-m^2)\tl + g(\omg, \tl) =0\, . 
\ee 
This is equivalent to
the equation \refb{e2.9} in string field theory. Of course, in order to
show that the perturbed theory is conformal we must ensure that the
$\beta$-functions associated with all other operators also vanish. To this 
end
we note that if $O$ denotes a boundary operator of dimension $h_O$, and if
we add this operator with coefficient $\lambda_O$ in the perturbed theory,
then the $\beta$-function of the coupling $\lambda_O$ is given by 
\be \label{e4.4} 
\beta_O = (h_O -1) \lambda_O + g_O(\tl, \lambda_O, \ldots)\, 
,
\ee 
where $\ldots$ denote the coupling constants associated with the other
operators added to the theory, and $g_O$ is the contribution to $\beta_O$
from higher order quantum corrections. Since $g_O$ is of order $\tl^2$ or
higher, as long as $(h_O-1)$ is of order 1, we can make \refb{e4.4} vanish
by choosing a $\lambda_O$ of order $\tl^2$. This procedure breaks down if 
$(h_O-1)$ is of order $\tl$ or higher for some operator $O$ other than 
$\cos(\omg X) V_T$ (which has already been discussed in eq.\refb{e4.3}). 
Thus we require that the operator product of $\cos(\omg X) V_T$ with 
itself does not generate an operator of dimension $1+\OO(\tl)$ other
than $\cos(\omg X) V_T$. 
This is 
identical to the condition derived in the previous section for 
obtaining 
a solution of the string field theory equations of motion in a 
perturbation expansion in $\tl$.

We shall now describe explicit computation of $\omg$ to order $\tl^2$.
For this we can set $\omg=m$ in the second term in \refb{e4.3}. This 
gives
\be \label{egive}
\omg^2 = m^2 - \tl^{-1} g(m,\tl)\, .
\ee
Thus
we need to compute $g(m,\tl)$.
For this computation we can pretend 
that the $X$-coordinate is
compactified with period $2\pi/m$. In that case $\cos(mX)V_T$, being a
dimension 1 operator, represents a massless scalar field $\vp$ in this
auxiliary
theory. Let $\VV(\vp)$ denote the tree level effective potential for this
scalar field 
obtained after integrating out all the other fields. 
(For $\omg\ne m$, there is also a 
`mass term' ${1\over 2} (\omg^2-m^2)\vp^2$ coming from the 
dimension of 
the operator $\cos(\omg X)V_T$.)
Then
by the usual correspondence between the
equations of motion of
string theory and the condition of conformal invariance, we can compute
the $\beta$-function $g(m,\tl)$ as
\be \label{e5.4}
g(m,\tl) = \VV'(\tl)\, ,
\ee
for suitably normalized $\VV$.
In order to compute $\VV(\vp)$ we simply compute the tree level amplitude
involving the massless scalars $\vp$ in this auxiliary string theory.
First of
all it is clear that there is no $\vp^3$ term in $\VV(\vp)$, since the
three
point amplitude of the scalar field described by the vertex operator
$\cos(mX) V_T$ vanishes due to $X$-momentum conservation. Thus the leading
contribution to $\VV(\vp)$ comes at the quartic order. The coefficient of 
this term can be
identified as $-(\NN^{-1}/4!)$ times the on-shell four point amplitude 
$A^{(4)}$ 
of the
auxiliary scalar
field at zero momentum:
\be \label{e5.4a}
\VV(\vp) = -(\NN^{-1}/4!) \, A^{(4)} \, \vp^4 + \OO(\vp^6)\, ,
\ee
where $\NN$ is the BPZ norm of the state $\cos(mX(0))V_T(0)|0\ra$ in the 
matter sector.\footnote{The simplest way to keep track of the 
normalization factor 
$\NN$ is to first do the computation by assuming $\cos(mX)V_T$ to be a 
normalized operator, and then account for the normalization by 
replacing $A^{(4)}$ by $\NN^{-2} A^{(4)}$ and $\tl$ by $\NN^{1/2}\tl$ in 
\refb{egg2}.}
Thus we get, from \refb{e5.4},
\be \label{egg1}
g(m,\tl) = -(\NN^{-1}/6) \, A^{(4)} \tl^3 + \OO(\tl^5)\, ,
\ee 
and hence from \refb{egive}
\be \label{egg2}
\omg^2 = m^2 + {\NN^{-1}\over 6} \, A^{(4)}\, \tl^2 + \OO(\tl^4)\, .
\ee
This of course agrees with eq.\refb{eexp8} derived from string field 
theory.

At this stage we should mention one subtle point. 
Interpreting 
the perturbed theory \refb{e4.1} as a deformed BCFT in the sense of
renormalization group flow makes sense strictly 
if the operator $\cos(\omg X) V_T$ is a relevant perturbation, {\i.e.} if 
$\omg^2< m^2$. {}From \refb{e4.3}, \refb{e5.4} we see that this requires 
$g(\omg, 
\tl)$, or equivalently $\VV(\vp)$, to be positive near $\vp=0$. In the 
analysis of 
string field theory in section 
\refb{s3} we did not encounter such a constraint. If $\VV$ turns out to be 
negative, we can still find a solution in string field theory by choosing 
$\omg^2-m^2$ to be positive. This will correspond to a net potential for 
the auxiliary scalar field
$\vp$ which has a minimum at $\vp=0$ and a maximum at a nearby point, and 
the non-trivial solution of \refb{e4.3} will correspond to the maximum of 
the potential. Thus even for negative $\VV$, we should expect a family of
BCFT labelled by $\tl$, although members of this family are not obtained
as deformation of the original BCFT by a relevant perturbation. Rather,
the original BCFT is obtained as a relevant deformation of each member of
this family.

When $\omega^2< m^2$ so that the perturbing operator is relevant, it is
more natural to treat $\omega$ rather than $\lambda$ as the independent
parameter. We simply deform the theory by the operator $\cos(\omega X)
V_T$, and consider the boundary CFT to which the theory flows in the
infrared. This will give a family of boundary CFT's labeled by $\omega$.
Inverse Wick rotation of these boundary CFT's then give us the family of
time dependent solutions corresponding to different initial conditions.

As special examples of the general class of perturbed BCFT's discussed 
above, we can consider the case where
$g(m, \tl)$ vanishes for arbitrary $\tl$, and as a 
result \refb{e4.3} is satisfied for $\omg=m$ for all $\tl$. In this case 
the 
operator $\cos(mX)\, V_T$ is an exactly marginal operator. The cases 
analysed in refs.\cite{0203211,0203265} are of this type.

Given a BCFT obtained by perturbing the euclidean theory by the operator
\refb{e4.1}, we can construct a BCFT by the inverse Wick rotation of this
perturbed theory. This will correspond to adding a perturbation
\be \label{e4.5}
\tl \cosh(\omg X^0(0)) \, V_T(0)\, c_1|0\ra\, ,
\ee
to the world-sheet theory with Minkowski signature space-time. The 
boundary state\cite{boundary} associated with this BCFT, containing
information about the 
time evolution of the energy-momentum tensor\cite{em-tensor,0203265}, can
be
obtained from Wick 
rotation of the boundary state associated with the euclidean theory. Time
evolution of the sources of other massless closed string fields, {\it
e.g.} the dilaton, the anti-symmetric tensor field $B_{\mu\nu}$, and
various Ramond-Ramond fields, can also be extracted from the boundary
state.

To see more explicitly how we can extract the energy-momentum tensor from
the boundary state, we recall that\cite{0203265} if part of the
boundary state 
involving oscillators of level (1,1) acting on the ghost number 3 vacuum 
has the form
\be \label{e5.17}
\int d^{26} k \, [\wt A_{\mu\nu}(k) \alpha^\mu_{-1}
\bar\alpha^\nu_{-1} 
+ \wt B(k) (\bar
b_{-1}
c_{-1} + b_{-1} \bar c_{-1})] (c_0+\bar c_0)c_1\bar c_1 |k\ra\, ,
\ee
then the energy momentum tensor $T_{\mu\nu}$ is given by:
\be \label{e5.18}
T_{\mu\nu}(x) = K (A_{\mu\nu}(x) + \eta_{\mu\nu} B(x) )\, ,
\ee
where $A_{\mu\nu}(x)$ and $B(x)$ are Fourier transforms of $\wt
A_{\mu\nu}(k)$ and $\wt B(k)$ respectively, and $K$ is a $\tl$-independent
constant. 
In the Wick rotated theory the energy momentum tensor computed using this
procedure will have the form:
\be \label{eem1}
T_{\mu\nu}(x, \vec x) = \sum_{n=0}^\infty T^{(n)}_{\mu\nu}(\vec x)
\cos(n\omega x)\, ,
\ee
where $\vec x$ denotes the spatial coordinates of the original theory.
Then after Wick rotation we have:
\ben \label{eem2}
T_{00}(x^0, \vec x) &=& - \sum_{n=0}^\infty T^{(n)}_{xx}(\vec
x)\cosh(n\omega
x^0)\, , \nonumber \\
T_{0i}(x^0, \vec x) &=& i \sum_{n=0}^\infty T^{(n)}_{xi}(\vec 
x)\cosh(n\omega x^0)\, , \nonumber \\
T_{ij}  &=& \sum_{n=0}^\infty T^{(n)}_{ij}(\vec
x)\cosh(n\omega x^0)\, .
\een
Since in the Wick rotated theory the sum over $n$ in \refb{eem1} is
expected to converge
and generate a finite energy-momentum tensor, we expect that after Wick
rotation the sum over $n$ in \refb{eem2} will converge for sufficiently
small $x^0$ and $\tl$. We
do expect the sum to diverge for large $x^0$, and checking
whether the results for small $x^0$ can be analytically continued beyond
the range of convergence or not, we shall have to determine whether during 
the time
evolution the energy momentum tensor hits a real singularity at a
finite value of $x^0$. 

In principle, we should also be able to extract the energy-momentum tensor
from the solution of the string field theory equations of motion
constructed in the previous section, since this solution is expected to
have all the information about the deformed boundary CFT. At present
however it is not known how to do this.

\sectiono{Explicit Example} \label{s5}

In this section we shall discuss an explicit example of the construction 
described in the previous section. The system that we shall consider is a 
D-$p$-brane of bosonic string theory, with one direction wrapped on a 
circle of radius $R>1$. 
This has the usual tachyonic mode 
of mass$^2=-1$, whose time evolution was discussed in \cite{0203211}. But 
this also has another tachyonic mode of mass$^2$
\be \label{em.1}
R^{-2}-1\equiv -m^2
\ee
coming from 
the 
first momentum mode of the standard tachyon along the circle. If we denote 
by $y$ the coordinate along the circle, then the matter sector vertex
operator 
associated with this tachyonic mode (with zero momentum along the
non-compact directions) can be taken to be:
\be \label{e5.1}
V_T = \cos(Y/R)\, .
\ee
We shall discuss the rolling of this tachyon 
when it is displaced from its maximum. As usual we denote by $x$ the Wick 
rotated time coordinate. The conformal field theory 
associated with the 
(Wick rotated) rolling tachyon solution is obtained by perturbing the 
original free conformal field theory by the boundary operator
\be \label{e5.2}
\tl\,  \int dt \, \cos(\omg X(t)) \, \cos(Y(t)/R)\, .
\ee
Eq.\refb{e4.3} now takes the form:
\be \label{e5.3}
(\omg^2 + R^{-2} - 1)\tl + g(\omg,\tl) = 0\, .
\ee
To leading order the solution to this equation is 
$\omg=m=\sqrt{1-R^{-2}}$. 
{}From eq.\refb{egg2} we see that the computation of $\OO(\tl^2)$ 
correction to $\omega$ 
requires us to compute $A^{(4)}$ involving four external states associated 
with the matter sector vertex operator $\cos(m X) \cos(Y/R)$.
This is the task to which 
we shall now turn.

Expressing the scalar field vertex operator as
\be \label{e5.5}
\cos(mX) \cos(Y/R) = {1\over 4} (e^{imX}+e^{-imX}) \, 
(e^{iY/R} + e^{-iY/R})\, ,
\ee
we can relate the four point amplitude $A^{(4)}$ of the scalar field to 
the 
Veneziano amplitude involving external states associated with 
matter sector vertex operators $e^{\pm i m X} e^{\pm i Y/R}$.
Since the right hand side of eq.\refb{e5.5} has four terms, there are
altogether $4^4=256$ terms involved in the computation of the four point
amplitude $A^{(4)}$. However all but 36 of them
vanish by $X$- or $Y$-momentum conservation. 12 of the non-vanishing terms
are each given by ${1\over 4^4} V(s=-4, t=0)$, and each of the rest 24
terms is given by ${1\over 4^4} V(s=-4R^{-2}, t = -4(1 - R^{-2}))$, where
$V(s,t)$ is the Veneziano amplitude
\ben \label{e5.6}
V(s,t) &=& 2\, \bigg( {\Gamma(-1-s) \Gamma(-1-t)\over \Gamma(-2-s-t)}
+ {\Gamma(-1-s) \Gamma(-1-u)\over \Gamma(-2-s-u)}
+{\Gamma(-1-u) \Gamma(-1-t)\over \Gamma(-2-u-t)} \bigg), \nonumber \\ \cr
&& \qquad \qquad u=-4-s-t\, .
\een
Thus we get
\ben \label{e5.7}
A^{(4)} &=& {1\over 4^4} \, ( 12 V(s=-4, t=0) + 24 V(s=-4 R^{-2}, 
t=-4(1-R^{-2})) \nonumber \\
&=& -{3\over 32} \, (R^2 - 2) [ 4 R^{-2} \ff'(4R^{-2}) / \ff(4R^{-2}) + 
\ff(4 
R^{-2}) - 1]\, ,
\een
where
\be \label{e5.8}
\ff(x) = \Gamma(1-x) \, \Gamma(1+x) = {\pi x\over \sin(\pi x)}\, .
\ee
Using eqs.\refb{egg2}, \refb{e5.7}, and the fact that $\NN=1/4$ for the
operator \refb{e5.5}, we now get
\be \label{e5.10}
\omg^2 = 1 - R^{-2} - {1\over 16} (R^2-2)  [ 4 R^{-2} \ff'(4R^{-2}) / 
\ff(4R^{-2}) + \ff(4
R^{-2}) - 1]\tl^2 + \OO(\tl^4)\, .
\ee
This determines the value of $\omg$ to order $\tl^2$ for which 
perturbation by the vertex 
operator $\cos(\omg X) \cos(Y/R)$ generates a boundary CFT. The time 
dependent solution describing the rolling tachyon is then 
determined by the inverse Wick rotation $X=-iX^0$ of this solution. In 
particular by inverse Wick rotating the boundary state associated with the 
perturbed euclidean BCFT, we can determine the energy-momentum tensor 
associated with the time dependent solution.

Unfortunately for generic $R$ the perturbed BCFT is not solvable, and 
hence 
we cannot explicitly compute the boundary state associated with this BCFT. 
For $R\to\infty$ the amplitude given in eq.\refb{e5.7} vanishes. This is 
consistent with the fact that in this limit the perturbing operator 
$\cos(X)$ is exactly marginal and hence the corresponding $\beta$-function 
must vanish. The associated conformal field theory has been
discussed in detail in \cite{marginal} and its inverse Wick rotation has
been used
in \cite{0203211,0203265} to describe the rolling of a spatially
homogeneous tachyon field on the D-brane. 
$A^{(4)}$ also vanishes at $R=1$ where the perturbing
operator is $\cos(Y)$. There is another value of $R$
where $A^{(4)}$ vanishes, 
namely at $R=\sqrt 2$.\footnote{Note that $\ff(4R^{-2})$ diverges at 
$R=\sqrt 2$ and so in evaluating \refb{e5.10} for $R=\sqrt{2}$ we need to 
carefully take the limit. Nevertheless the right hand side of \refb{e5.10} 
can be shown to vanish at $R=\sqrt 2$.}
Thus one might expect that the corresponding perturbation becomes exactly 
marginal at this point. To see that this is indeed the case, let us note 
that at $\omega=1/\sqrt 2$, the perturbing operator is
given by:
\be \label{e5.11}
\tl \, \int dt \, \cos(X(t)/\sqrt 2) \, \cos(Y(t)/\sqrt 2)
= {\tl\over 2} \, \Big[\cos\Big((X(t) + Y(t))/\sqrt 2\Big) + 
\cos\Big((X(t) 
-
Y(t))/\sqrt 2\Big)\Big]\, .
\ee
This corresponds to perturbing the theory by a sum of two commuting 
exactly marginal 
deformations, and hence represents a marginal deformation. The associated 
BCFT is exactly solvable. In fact, defining
\be \label{e5.12}
Z^1 = (X+Y)/\sqrt 2 \, , \qquad Z^2 = (X-Y)/\sqrt 2\, ,
\ee
we see that the net boundary state is given by the product
\be \label{e5.13}
|\BB\ra_{Z^1} \otimes |\BB\ra_{Z^2} \otimes |\BB\ra_{c=24} \otimes 
|\BB\ra_{ghost}\, .
\ee
$|\BB\ra_{c=24}$ denotes the boundary state associated with the 24 free 
scalar fields, and $|\BB\ra_{ghost}$ denotes the boundary state associated 
with the ghost fields. These are given by their free field expressions. On 
the 
other hand the relevant part of the boundary states $|\BB\ra_{Z^1}$ and $ 
|\BB\ra_{Z^2}$, associated with the perturbed BCFT associated with the 
scalar fields $Z^1$ and $Z^2$, can be read out from the results of 
\cite{marginal,9811237,0108238,0203265}. If we consider for simplicity the
case of 
the 
D-25-brane, the part of the boundary state, relevant for the computation 
of the energy-momentum tensor, is proportional to:
\ben \label{e5.14}
&& \bigg[f\Big((iX(0)+iY(0))/\sqrt 
2\Big) 
- {1\over 2} (\alpha^X_{-1} + \alpha^Y_{-1}) (\bar\alpha^X_{-1} + 
\bar\alpha^Y_{-1}) g\Big((iX(0)+iY(0))/\sqrt 2\Big) \bigg] \nonumber \\
&& \bigg[f\Big((iX(0)-iY(0))/\sqrt 2\Big)
- {1\over 2} (\alpha^X_{-1} - \alpha^Y_{-1}) (\bar\alpha^X_{-1} -
\bar\alpha^Y_{-1}) g\Big((iX(0)-iY(0))/\sqrt 2\Big) \bigg] \nonumber \\
&& \bigg[1-\alpha^i_{-1}\bar\alpha^i_{-1}\bigg]  \, \bigg[1-\bar 
b_{-1}
c_{-1} - b_{-1} \bar c_{-1}\bigg] (c_0+\bar c_0)c_1\bar c_1 |0\ra\, .
\een
Here $\alpha^i_{-1}$, $\bar\alpha^i_{-1}$ for $2\le 
i\le 25$ are the 
oscillators associated with the spectator bosons $X^i$,
$\alpha^X_{-1}$, $\bar\alpha^X_{-1}$, $\alpha^Y_{-1}$, $\bar\alpha^Y_{-1}$ 
are the oscillators associated with the scalars $X$ and $Y$, and $b_n$,
$c_n$, $\bar b_n$, $\bar c_n$ are the oscillators associated with the
ghost fields $b$, $c$, $\bar b$, $\bar c$.
The functions $g(x)$ and $f(x)$ have been given in \cite{0203211,0203265}:
\be \label{e5.15}
f(x^0)={1\over 1 + e^{x^0} \sin(\tl\pi/2)} + {1 \over
1 + e^{-x^0} \sin(\tl\pi/2)} - 1\, , \qquad
g(x^0) = \cos(\tl\pi) +1 - f(x^0)\, .
\ee
Note that these formul\ae\ differ from that of \cite{0203211,0203265} by a 
replacement $\tl\to \tl/2$, whose origin can be traced to the explicit 
factor of $1/2$ multiplying $\tl$ on the right hand side of \refb{e5.11}.

We can now make an inverse Wick rotation $X=-iX^0$ to get the boundary 
state in the Minkowski signature theory:
\ben \label{e5.16}
&& \bigg[f\Big((X^0(0)+iY(0))/\sqrt 
2\Big) 
- {1\over 2} (-i\alpha^0_{-1} + \alpha^Y_{-1}) (-i\bar\alpha^0_{-1} + 
\bar\alpha^Y_{-1}) g\Big((X^0(0)+iY(0))/\sqrt 2\Big) \bigg] \nonumber \\
&& \bigg[f\Big((X^0(0)-iY(0))/\sqrt 2\Big)
- {1\over 2} (-i\alpha^0_{-1} - \alpha^Y_{-1}) (-i\bar\alpha^0_{-1} -
\bar\alpha^Y_{-1}) g\Big((X^0(0)-iY(0))/\sqrt 2\Big) \bigg] \nonumber \\
&& \bigg[1-\alpha^i_{-1}\bar\alpha^i_{-1}\bigg]  \, \bigg[1-\bar 
b_{-1}
c_{-1} - b_{-1} \bar c_{-1}\bigg] (c_0+\bar c_0)c_1\bar c_1 |0\ra\, .
\een
Comparing \refb{e5.16} with \refb{e5.17} we get,
\ben \label{e5.19}
A_{00} &=& {1\over 2} \, [g((x^0 + iy)/\sqrt 2) f((x^0 - i
y)/\sqrt
2)  + g((x^0 - i 
y)/\sqrt 
2) f ((x^0 + iy)/\sqrt 2)]\, , \nonumber 
\\
A_{0y} &=& A_{y0} = {i\over 2}  [g((x^0 + iy)/\sqrt 2) f((x^0 - i 
y)/\sqrt 
2)  - 
g((x^0 - i 
y)/\sqrt
2)  f ((x^0 + iy)/\sqrt 2)]\, , \nonumber
\\
A_{yy} &=& -  {1\over 2} \,  [g((x^0 + iy)/\sqrt 2)f((x^0 - i
y)/\sqrt
2) + g((x^0 - 
i 
y)/\sqrt
2)  f ((x^0 + iy)/\sqrt 2) ]\, , \nonumber
\\
A_{ij} &=& - \, \delta_{ij} \, 
f ((x^0 + iy)/\sqrt 2)  f((x^0 - i
y)/\sqrt
2)
\, , \nonumber
\\
B &=& -  f ((x^0 + iy)/\sqrt 2)  f((x^0 - i
y)/\sqrt
2)\, .
\een
Using \refb{e5.18}, \refb{e5.15} and \refb{e5.19} we get
\ben \label{e5.20}
T_{00} &=& K (A_{00} - B) = {K\over 2} (\cos(\tl\pi)+1) [f ((x^0 + 
iy)/\sqrt 
2) + f((x^0 - i
y)/\sqrt
2)]\, , \nonumber \\
T_{0y} &=& K A_{0y} = {i K\over 2}  (\cos(\tl\pi)+1) [f((x^0 - i
y)/\sqrt
2) - f ((x^0 +
iy)/\sqrt
2) ] \nonumber \\
T_{yy} &=&  K (A_{yy} + B) = -{K\over 2} (\cos(\tl\pi)+1) [f ((x^0 +
iy)/\sqrt
2) + f((x^0 - i
y)/\sqrt
2)]\, , \nonumber \\  
T_{ij} &=& K(A_{ij} + B \delta_{ij})
=  - 2 K\, \delta_{ij} \, f ((x^0 +
iy)/\sqrt
2) \, f((x^0 - i
y)/\sqrt
2)
\een
All other components of $T_{\mu\nu}$ vanish.
{}From these expressions one can easily verify the conservation laws:
\be \label{e5.21}
\p_0 T_{00} - \p_y T_{0y} = 0, \qquad \p_0 T_{0y} - \p_y T_{yy} = 0\, .
\ee
If we consider a general D-$p$-brane with $p<25$, then $T_{ij}$ associated
with the $(25-p)$ transverse directions vanish. Furthermore the
non-vanishing components of $T_{\mu\nu}$ are each multiplied by a delta
function involving the transverse coordinates. The $\tl$ independent
constant $K$ is computed by requiring that for $\tl=0$, $T_{00}$ should
reproduce the tension $\TT_p$ of a D-$p$-brane. This gives
\be \label{ektp}
K = {\TT_p\over 2}\, .
\ee

{}From \refb{e5.15} we get
\be \label{e5.22}
f((x^0\pm i y)/\sqrt 2) = 
{1\over 1 + e^{x^0/\sqrt 2} e^{\pm iy/\sqrt 2} \sin(\tl\pi/2)} + {1 \over
1 + e^{-x^0/\sqrt 2}  e^{\mp iy/\sqrt 2} \sin(\tl\pi/2)} - 1\, .
\ee
Thus $f((x^0\pm iy)/\sqrt 2)$ and hence
the energy momentum tensor hits a singularity at
\be \label{e5.23}
x^0 = \sqrt 2 \, \ln \bigg|{1\over \sin(\tl\pi/2)}\bigg| \equiv x^0_c,
\qquad y = \cases {\sqrt 2 \pi\, , \qquad \hbox{for} \quad \tl>0\, , \cr
0\, , \qquad \, \, \quad \, \hbox{for} \quad \tl<0\, .}
\ee
In particular the energy-density at the singular point approaches 
$\infty$. Thus we cannot naively evolve the system beyond this time using
the 
classical open string field equations.

\begin{figure}[!ht]
\leavevmode
\begin{center}
\epsfysize=8cm
\epsfbox{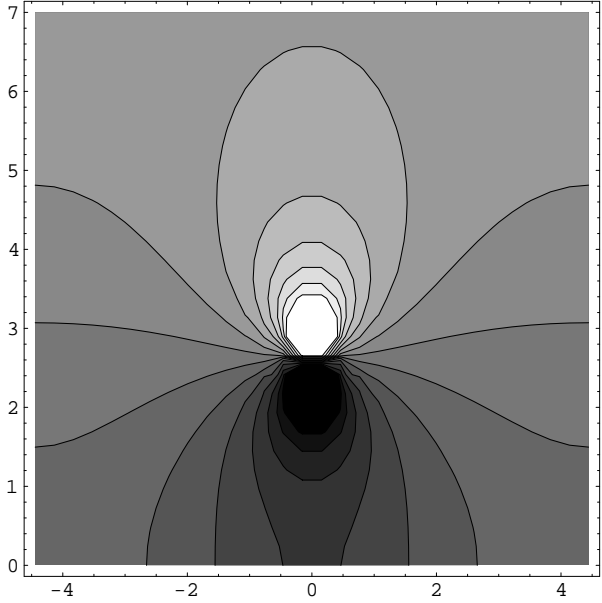}
\end{center}
\caption{The contour plot of the time evolution of the energy density
$T_{00}$ computed from eq.\refb{e5.20}, \refb{e5.22} for $\tl=-0.1$. The
horizontal axis
is the $y$ axis, the vertical axis is the time ($x^0$) axis, and the
shading at a
given point $(y,t)$ shows the value of $T_{00}$, with darker shade
representing higher energy density. We see from this that the $T_{00}$
hits a singularity at $x^0\simeq 2.62$, $y=0$. At this space-time point, 
the system
loses a net amount of energy equal to the total initial energy of the
system.  Immediately after this
singularity the energy density becomes negative around $y=0$, as shown by 
the lighter regions, and eventually the energy density approaches zero
everywhere.} \label{f2} \end{figure}
It is instructive to examine the nature of the singularity in a little 
more detail. Using
eqs.\refb{e5.20}, \refb{e5.22} we get the total energy of the system to 
be:
\be \label{etote}
\EE = \int_0^{2\pi\sqrt 2} dy \, T_{00} = 2\pi\sqrt 2 K 
(1 + \cos(\tl\pi))\, ,
\ee
for $x^0< x^0_c$. If we
naively 
continue the expression for $T_{00}$ to $x^0>x^0_c$, we find 
that the total energy vanishes. In order to see the origin of this 
`violation' of energy conservation, we can examine the form of the 
singularity at $x^0 = x^0_c$, 
$y=0$ (concentrating for definiteness on the $\tl<0$ case.) Defining
\be \label{defth}
\Theta \equiv e^{(x^0 - x^0_c)/\sqrt 2} = -e^{x^0/\sqrt 2}
\sin(\tl\pi)\simeq 1
\ee
near the singularity, the singular part of $T_{00}$ near $(1-\Theta)\sim y
\simeq 0$ is given 
by:
\be \label{esingpart}
(T_{00})_{singular} = K(1 + \cos(\tl\pi)) \, {(1-\Theta) \over 
(1-\Theta)^2 + {1\over 2} y^2} \simeq \, {\rm sgn}(1-\Theta) \, \pi
\sqrt{2} \, K
(1 + \cos(\tl\pi)) \, 
\delta(y)\, .
\ee
Since $(1-\Theta)$ changes sign as $x^0$ passes from below to above
$x^0_c$, there is a net loss of energy at $y=0$ at $x^0=x^0_c$,
The amount of energy lost at $y=0$ as $x^0$ passes from below $x^0_c$
to above $x^0_c$ is $2\pi \sqrt{2} \, K (1 + \cos(\tl\pi))$, -- same
as the total 
energy of the system. As a result
the average of the energy 
density in the rest of the system must vanish after the critical time
$x^0_c$, as has already been seen in explicit calculation. If we
continue the 
formula for $T_{00}$ beyond this critical time, the system eventually
evolves to a configuration with zero 
energy density everywhere.\footnote{Of course, near the singularity
$x^0=x^0_c$, $y=0$, the back reaction of closed string fields will be
important, and a realistic analysis of the dynamics must include this
effect.} There is an apparent violation of energy conservation at
$x^0 = x^0_c$, but
the most conservative interpretation of this is simply that we create a 
codimension 1 D-brane at $y=0$, with a delta function source of energy
that is not visible in the formula \refb{e5.20}. The total energy stored
at $y=0$ does not 
agree with the tension of a codimension 1 D-brane; in particular the ratio
of the total energy of the system to that of a codimension 1 D-brane is
given by:
\be \label{eratio}
{1\over \sqrt 2} (1 + \cos(\pi\tl))\, .
\ee
However the final state
could represent a 
state of the codimension 1 D-brane with non-trivial time dependent tachyon
configuration, with the excess (deficit) energy stored in the 
tachyon field. Fig. \ref{f2} shows the time evolution of $T_{00}$ given in
eqs.\refb{e5.20}, \refb{e5.22} for a specific value of $\tl$.

\sectiono{Time Dependent Solutions on D-branes in Superstring 
Theories} \label{s6}

The analysis of the previous sections can be easily generalized to the 
case of superstring theories. Let us consider a generic D-brane system 
with a tachyonic mode, and let the zero momentum tachyon vertex operator 
have the form $e^{-\phi} V^{(-1)}_T$ in the $-1$ picture\cite{FMS}, where
$\phi$ 
denotes the bosonized ghost field and $V^{(-1)}_T$ denotes the matter part 
of the vertex operator. If $V^{(-1)}_T$ has dimension $h<1/2$, then the 
tachyon has mass$^2=(h-{1\over 2})\equiv -m^2$ and at the linearized 
level the time dependent tachyon solution takes the form $\tl\cosh(m 
x^0)$. In the Wick rotated theory this corresponds to $\tl\cos(mx)$, and 
amounts to perturbing the BCFT by a term
\be \label{e6.1}
\tl \int dt\, d\theta \cos(m X(t,\theta)) V^{(-1)}_T(t,\theta)\, ,
\ee
where $t$ denotes the coordinate labelling the world-sheet boundary, 
$\theta$ is the fermionic coordinate on the superspace, and we 
consider $X$ and 
$V_T^{(-1)}$ as 
superfields. Although to lowest order in $\tl$ this represents a marginal 
deformation, in general once higher order corrections are taken into 
account this no longer represents a marginal deformation. We remedy this 
by modifying \refb{e6.1} to 
\be \label{e6.2}
\tl \int dt \, d\theta \cos(\omg X(t,\theta)) V^{(-1)}_T(t,\theta)\, ,
\ee
where $\omg$ is a constant. The full $\beta$-function associated with 
the coupling $\tl$ now takes the form:
\be \label{e6.3}
\omg_{\tl} = (\omg^2 - m^2)\tl + g(\omg,\tl)\, ,
\ee
as in \refb{e4.2}. We determine $\omg$ for a given $\tl$ by 
demanding that $\beta_{\tl}$ vanishes. This gives:
\be \label{e6.4}
(\omg^2 - m^2)\tl + g(\omg,\tl) = 0\, .
\ee
This determines $\omg$ as a function of $\tl$. Thus for example the 
leading correction to $\omg$ is given by
\be \label{el1}
\omg^2 \simeq m^2 - g(m, \tl) / \tl\, .
\ee
The $\beta$-functions associated with the other operators can be made to 
vanish as in section \ref{s4} provided the product of \refb{e6.2} 
with itself does not generate integral of any 
superfield of dimension $1/2+\OO(\tl)$ other than the superfield 
$\cos(\omg
X(t,\theta)) V^{(-1)}_T(t,\theta)$ itself.

For an explicit example we can consider the case analogous to the one 
discussed in section \ref{s5} for open bosonic string theory. We take a 
non-BPS D-$p$-brane with one direction compactified on a circle of radius 
$> {\sqrt 2}$. Let $y$ denote this compact direction. In this case the 
first momentum mode of the tachyon along the circle, described by the 
wave-function $\cos(y/R)$, can be thought of as a scalar field living
in $((p-1)+1)$ 
dimensions with mass$^2$:
\be \label{e6.5}
{1\over R^2} - {1\over 2} \equiv -m^2\, .
\ee
Displacing this tachyonic mode from its maximum and letting it roll 
amounts to displacing the full $(p+1)$ dimensional tachyon from its 
maximum by an amount proportional to $\cos(y/R)$ and let it roll. Since 
the sign of the displacement is different for different values of $y$, we 
expect that the time evolution of the tachyon will produce a kink-antikink 
pair on the D-$p$-brane world volume, centered around $y=\pm\pi R/2$.

We shall study the time evolution by viewing this particular mode of the 
tachyon as a scalar field in $((p-1)+1)$ dimensions.
The vertex operator of this zero momentum tachyon in the $-1$ picture is 
$e^{-\phi}\cos(Y/R)\otimes \sigma_1$
where $\sigma_1$ is the Chan-Paton factor\cite{9808141,9809111}.
Thus in order to construct the Wick rotated version of the rolling tachyon 
solution, we need to perturb the BCFT by the operator
\be \label{e6.6}
\tl \int dt d\theta \cos(\omg X(t,\theta)) \cos(Y(t,\theta)/R)\otimes 
\sigma_1\, .
\ee
If we expand the superfields $X(t,\theta)$ and $Y(t,\theta)$ as
\be \label{e6.7}
X(t,\theta) = X(t) + \theta\psi_x(t), \qquad
Y(t,\theta) = Y(t) + \theta\psi_y(t),
\ee
where $\psi_x$ and $\psi_y$ are the world-sheet fermionic partners of the 
$X$ and the $Y$ fields respectively, 
then the perturbation can be expressed as
\be \label{e6.8}
-\tl \int dt [\omg \sin(\omg X(t)) \cos(Y(t)/R) + R^{-1} 
\cos(\omg X(t)) \sin(Y(t)/R)]\otimes \sigma_1\, .
\ee
As in section \ref{s5}
we can relate the computation of $g(m,\tl)$ appearing in 
eq. \refb{el1} to leading order in $\tl$
to the computation of the four point amplitude of four 
on-shell external states, each described the the vertex operator 
$e^{-\phi} \cos(m X) \cos(Y/R)\otimes \sigma_1$. This in turn will 
determine the leading 
correction to $\omg$ via eq.\refb{el1}. 
We shall not carry out the detailed computation here. 

We can consider the special value of the radius $R=2$. In this case 
$m=1/2$, and to leading order when we set $\omg=m$ the perturbation 
\refb{e6.6} takes the form:
\ben \label{e6.9}
&&\tl \int dt d\theta \cos(X(t,\theta)/2) \cos(Y(t,\theta)/2)\otimes 
\sigma_1 \nonumber \\
&=& {\tl\over 2} \int dt \, d\theta \, [ \cos(Z_1(t,\theta)/\sqrt 2) + 
\cos(Z_2(t,\theta)/\sqrt 2) 
]\otimes \sigma_1\, ,
\een
where
\be \label{e6.10}
Z_1 = (X+Y)/\sqrt 2, \qquad Z_2 = (X-Y)/\sqrt 2\, .
\ee
Writing
\be \label{e6.11}
Z_i(t,\theta) = Z_i(t) + \theta \psi_i(t)\, ,
\ee
we can rewrite the perturbation as
\be \label{e6.12}
-{\tl\over 2\sqrt 2} \int dt [\psi_1(t) \sin (Z_1(t)/\sqrt 2)
+ \psi_2(t) \sin (Z_2(t)/\sqrt 2)]\otimes \sigma_1\, .
\ee
Following the results of \cite{9808141,0003124} one can show that each of 
the two terms 
in \refb{e6.12} represents an exactly marginal deformation. Unfortunately 
however 
these two terms do not commute, instead they anti-commute due to the 
presence of the fermion fields $\psi_i$. As a result the complete 
perturbation is not exactly marginal and the $\beta$-function does receive 
higher order corrections. Thus unlike in the case of section \ref{s5}, we 
cannot solve the theory exactly in this case.
In particular,
we cannot obtain an analytic expression for the 
boundary state and compute the energy momentum tensor associated with 
this solution.

To see this more explicitly we fermionize  and rebosonize following the 
procedure given in refs:\cite{9808141,0003124}. On the boundary the 
relevant part of these 
relations take the form:
\be \label{e6.13}
e^{iZ_i(t)/\sqrt 2} = {1\over \sqrt 2} (\xi^i(t) + i \eta^i(t)) \otimes 
\tau_i\, ,
\ee
\be \label{e6.15}
\psi_i(t) \eta_i(t) = -{i\over \sqrt 2} \p \phi_i(t)\, ,
\ee
where $\xi_i$ and $\eta_i$ are Majorana fermions, $\phi_i$ are scalars and 
$\tau_i$ are Pauli matrices providing the cocycles needed for 
the bosonization. In terms of the bosonic field $\phi_i$ the perturbation 
\refb{e6.12} can be written as:
\be \label{e6.16}
{i\tl\over 4\sqrt 2} \int dt [\p_t\phi_1 \otimes \tau_1\otimes \sigma_1
+ \p_t\phi_2 \otimes \tau_2 \otimes \sigma_1]\, .
\ee
Since $\tau_1\otimes \sigma_1$ does not commute with 
$\tau_2\otimes\sigma_1$, the above perturbation can be thought of as 
switching on a pair of non-commuting Wilson lines, one along $\phi_1$ and 
the other along $\phi_2$, each being of magnitude $\tl/4\sqrt 2$. As a 
result the 
potential does not vanish. Instead we get a contribution proportional to
the square of the commutator, which in this case will be equal to $C\tl^4$ 
for 
an approprite positive constant $C$. This gives a contribution to 
$g(m,\tl)$ equal 
to $4 C\tl^3$. Eq.\refb{el1} then gives:
\be \label{e6.17}
\omg^2 = {1\over 4} - 4 C \tl^2\, .
\ee
This correction to the value of $\omg$ shows that \refb{e6.9} does not 
represent an exactly marginal deformation.

If we had naively ignored the issue of non-commutativity of the two 
operators in \refb{e6.16} and treated this as an exactly marginal 
deformation, then we would end up with an expression for the energy 
momentum tensor similar to that in \refb{e5.20}, with $f(x^0)$ given 
by\cite{0203265}
\be \label{e6.18}
f(x^0) =  {1\over 1 + e^{\sqrt 2 x^0} \sin^2(\tl\pi/2)} + {1 \over
1 + e^{-\sqrt 2 x^0} \sin^2(\tl\pi/2)} - 1\, .
\ee
{}From this we see that $f((x^0\pm iy)/\sqrt{2})$ has singularities at 
$x^0=\ln(1/\sin^2(\tl\pi/2))$, $y=\pm \pi$, and the energy density blows 
up at these space-time points. Thus energy flows into these points from 
the rest 
of the system. These are precisely the points where we expect the 
codimension 1 brane-antibrane pairs to form, and the nature of these 
singularities has precisely the same delta function form as discussed at 
the end of section \ref{s5}.\footnote{The analysis of RR charge density
shows that RR charge also accumulates at these points.} Thus one might be
tempted to interpret
this as the creation of brane-antibrane pair.\footnote{Production of
brane-antibrane pair from the decay of a higher dimensional brane has also
been discussed in a different approximation in \cite{0204203}.} However,
once we take into 
account the non-marginality of the perturbation \refb{e6.16} this result 
will be modified. In fact, if the perturbation \refb{e6.12} had been
exactly marginal, and the energy momentum tensor had been given by
\refb{e5.20}, \refb{e6.18}, we would have an inconsistency. To see this,
note that following arguments similar to those in section \ref{s5} we
would conclude in this case that the total initial energy of the system is
deposited into a codimension 1 brane-antibrane pair, each carrying an
energy density
equal to:
\be \label{e6.19}
{1\over \sqrt 2} \, \TT_{p-1} \, (1 + \cos(\tl\pi))\, ,
\ee
where $\TT_{p-1}$ is the tension of a BPS D-$(p-1)$ brane. Thus for a
generic $\tl$, \refb{e6.19} differs from the tension of a BPS
D-$(p-1)$-brane. 
Unlike in the case of bosonic D-branes, in this case we
cannot attribute the excess (deficit) energy to the energy stored in the
tachyon
field, since there are no tachyons on a BPS D-brane. (The excess energy
can still be attributed to other fields on the
BPS D-brane, but there is no way a brane can have energy density lower
than its tension.) Thus we would be led to an inconsistency, as there will
be no interpretation for the final state for $(1+\cos(\pi\tl))< \sqrt{2}$.
Thus
it is just as well that \refb{e5.20}, \refb{e6.18} does not give the
correct formula for the stress tensor.

\sectiono{Tachyon Dynamics in Closed String Theory} \label{s7}

In this section we briefly discuss possible extensions of the method 
outlined in this paper to the study of tachyon dynamics in closed string 
theory. 
Given a zero 
momentum
tachyon field described by a dimension $(h,h)$ vertex operator $V_T$, we 
try to construct a solution in the Wick rotated theory ($x^0\to i x$) by 
perturbing the orginal CFT by 
\be \label{e8.1}
\tl \int d^2 z \, \cos(\omg X(z,\bar z)) V_T(z, \bar z)\, .
\ee
For $\omg^2 = 4(1-h)$ this would describe a marginal deformation to
leading order in $\tl$. 
However since in general higher order contribution to the $\beta$-function 
does not vanish, we do not fix $\omg$ at the beginning. The 
$\beta$-function associated with the perturbation \refb{e8.1} then takes 
the form:
\be \label{e8.2}
\beta_{\tl} = 2\, \Big({1\over 4} \omg^2 + h - 1\Big)\, \tl + 
g(\omg,\tl)\, .
\ee
We now adjust $\omg$ for a given $\tl$ such that \refb{e8.2} vanishes. 
Thus this perturbation describes a conformal field theory.

The main obstruction to carrying out this program arises from the fact
that
in this case the operator product of \refb{e8.1} with itself does generate
other dimension (1,1) operators, -- namely the vertex operators associated
with the zero momentum graviton and dilaton fields.  (The source for the 
dilaton field manifests itself in the form of a change in the central 
charge induced by the perturbation.) Thus the rolling
tachyon soution acts as a source for the zero momentum graviton and the
dilaton fields, and {\it a priori} there is no guarantee that we shall be
able to solve the equations of motion for these fields perturbatively in
the parameter $\tl$. (See \cite{mukherji} for an analysis of this 
problem in non-polynomial closed string field theory\cite{non-pol}.) This
problem of course is not specific to string 
theory, and even in the case of a normal scalar field theory coupled to 
the 
graviton and the dilaton, there is no guarantee that given a time 
dependent solution involving the scalar field, we can solve for the 
graviton and the dilaton field perturbatively in the parameter labelling 
the solution of the scalar field equations. However, for tachyons 
localized on a 
subspace of the full space-time\cite{local}, it may be possible to solve 
the 
equations of motion of the graviton and the dilaton fields as a 
perturbation expansion in the deformation parameter.

\sectiono{Conclusion} \label{s8}

In this paper we have discussed a general method for constructing time
dependent solutions, describing the rolling of a tachyon on an unstable
D-brane system, in cubic open string field theory. We have also provided a
description of these solutions as Wick rotated version of euclidean
boundary conformal field theories. The family of time dependent solutions
associated with different initial conditions correspond to a family of
boundary conformal field theories, each of which is related to the
original D-brane system by a nearly marginal deformation. The construction
can be easily generalized to the case of unstable D-brane systems in
superstring theory.

In general these deformed boundary conformal field theories are not
exactly solvable, and hence we cannot find explicit analytic expressions
for the time dependence of the energy momentum tensor by Wick rotating the
boundary state associated with the boundary conformal field theory. 
However in some special cases the family of boundary CFT's are related by
an exactly marginal deformation and are exactly solvable. This allows us
to compute the time evolution of the energy momentum tensor explicitly for
arbitrary initial condition on the tachyon. One such example discussed in
this paper is the case of a bosonic D-$p$-brane wrapped on a circle of
radius $\sqrt 2$, and we consider the first momentum mode of the tachyon
along the circle. Pushing this particular tachyonic mode away from its
extremum corresponds to an initial condition where the tachyon field has
different signs in different parts of the circle, and hence we might
expect that the time evolution of the configuration may produce lower
dimensional D-branes. We explicitly construct the conserved energy
momentum tensor associated with this solution and show that the time
evolution of the system does indicate the creation of a codimension 1
brane. 

One would like to generalize this to describe creation of a lower
dimensional brane-antibrane pair due to the decay of an unstable D-brane
in superstring theory.
But unfortunately the situation here is more complicated as the associated
family of conformal field theories are not related to each other by
marginal deformation and are not exactly solvable. We hope to return to
this problem in the near future.

We conclude with two remarks:

\begin{itemize}
\item The solutions constructed here are obtained in {\it classical} open
string field theory where we ignore coupling to closed string fields. This
is expected to be a good approximation in the string coupling $g_s\to 0$
limit. Once we switch on $g_s$, two effects will have to be taken into
account in the study of the solution: classical backreaction of closed
string fields on the solution, and the possibility of the decay into
quantum closed string states. The time scale over which these effects will
affect the solution is expected to go to infinity in the $g_s\to 0$ limit,
but at present we do not know the precise dependence of this time scale on
$g_s$.

Another parameter that is important in determining the effect of closed
string coupling to the solution is $\tl$, or equivalently, the total
energy
density of the configuration. We would expect that for small energy
density, the effect of coupling to the closed strings will be small, but
again we do not know the precise dependence.

\item In the study of closed string tachyon condensation\cite{local},
renormalization group (RG) flow has played a useful role. 
The approach taken here is somewhat different: we interpret the complete
time dependent solution for a specific initial condition as a single
conformal field theory (after Wick rotation). Different inequivalent
initial conditions will generate different boundary conformal field
theories.

In fact it is possible to argue that at least for open string tachyon
condensation, RG flow can never describe classical time
evolution
in a strict sense. RG flow takes us from a BCFT with higher boundary
entropy to one of lower boundary entropy\cite{bent}. In the space-time
language this corresponds to interpolating between a D-brane of higher
energy to a D-brane of lower energy. It is clear that classical time
evolution can never do that; due to conservation of energy it must take us
from a D-brane of higher energy to one of lower energy plus other stuff.
As a result, the RG analysis can at most give some qualitative information
about time evolution in open string theory.

\end{itemize}

\medskip

{\bf Acknowledgement}:
I would like to thank R.~Brower, R.~Gopakumar, J.~Minahan, S.~Minwalla, 
N.~Moeller, L.~Rastelli and B.~Zwiebach for 
useful discussions, and B.~Zwiebach for critical comments on the
manuscript.
This work was supported in part by a grant 
from the Eberly College 
of Science of the Penn State University. I would also like to acknowledge 
the hospitality of the Center for Theoretical Physics at
MIT, and a grant from the NM Rothschild and Sons Ltd
at the Isaac Newton
Institute where part of this work 
was done.

\end{document}